\documentclass[sigconf,screen]{acmart}

\renewcommand\footnotetextcopyrightpermission[1]{} %
\usepackage{gensymb}

\usepackage{balance}

\newcommand{\etal}{\textit{et al. }}

\begin{document}

\title{OnionBot: A System for Collaborative Computational Cooking}

\author{Bennet Cobley}
\affiliation{\institution{Imperial College London}}
\email{bennet.cobley16@imperial.ac.uk}

\author{David Boyle}
\affiliation{\institution{Imperial College London}}
\email{david.boyle@imperial.ac.uk}

\begin{abstract}
An unsolved challenge in cooking automation is designing for shared kitchen workspaces. In particular, robots struggle with dexterity in the unstructured and dynamic kitchen environment. We propose that human-machine collaboration can be achieved without robotic manipulation. We describe a novel system design using computer vision to inform intelligent cooking interventions. This human-centered approach does not require actuators and promotes dynamic, natural collaboration. We show that automation that assists user-led actions can offer meaningful cooking assistance and can generate the image databases needed for fully autonomous robotic systems of the future. We provide an open source implementation of our work\footnote{An open source implementation of OnionBot is available at \href{https://github.com/onionbot}{github.com/onionbot}.} and encourage the research community to build upon it.
\end{abstract}



\begin{teaserfigure}
  \centering
  \includegraphics[width=\linewidth]{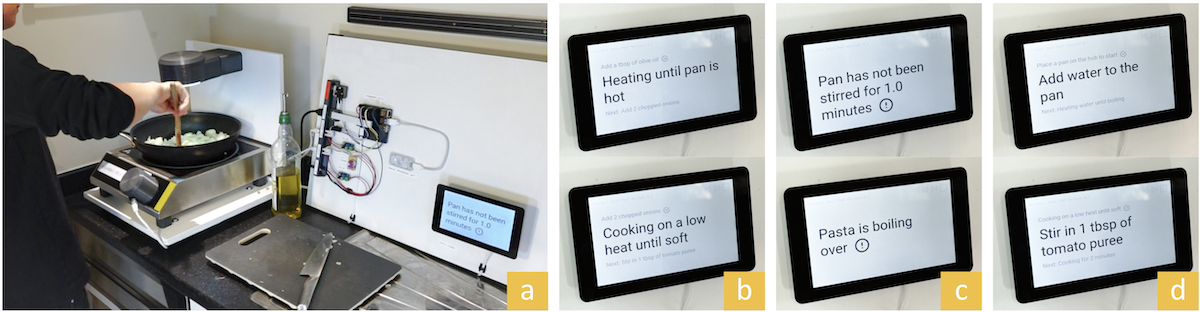}
  \label{abstractfigure}
  \caption{(a) Collaborative computational cooking. The system combines manual and automatic cooking assistance: (b) automatic management of heating tasks; (c) intelligent reminders and warnings; (d) on-screen instructions for the user to execute.}
  \Description{Section A: Photograph of a person stirring a frying pan on a stove on a kitchen counter. A nearby touchscreen says "Pan has not been stirred for 1 minute!" The stove has been modified with various components wired together. There is one module mounted where the control knob would be, and one module mounted above the pan, with LED lights shining into the pan. Next to the stove, a board mounts the screen and many components that are wired to the stove-mounted modules. Section B, C and D show pairs of different messages on a touch screen. Section B: Heating until pan is hot and Cooking on a low heat until soft. Section C: Pan has not been stirred for 1.0 minutes and Pasta is boiling over. Section D: Add water to the pan and Stir in 1 tbsp of tomato puree.}
\end{teaserfigure}
\maketitle

\pagestyle{empty}

\section{Introduction}
Automation devices have become an essential part of the home. Appliances that optimize specific functions take on repetitive cooking and cleaning tasks, bringing efficiency and convenience to domestic life. With the rapid development in ubiquitous computing and computer vision, we foresee that appliances will further assist with creative roles, including in the kitchen; \emph{computational cooking} will bring a new generation of partnership with food technology. 

Studies typically explore high degree-of-freedom robot arms as solutions for automated cooking \cite{bollini2013interpreting, junge2020improving, inagawa2020japanese, yi2020task, gravot2006cooking, zhai2012kinematic}. Robot arms mimic general human-kitchen interaction, but the dynamic and unstructured kitchen environment poses significant manipulation challenges. Consequently, most solutions compromise in a static, human-free workspace \cite{bollini2013interpreting, junge2020improving, inagawa2020japanese, yi2020task, gravot2006cooking, zhai2012kinematic}. 

Coexistence in a shared environment is critical to feasibility of next-generation systems. \textit{Collaborative robots} leverage human-robot interaction in shared workspaces, combining the best of human adaptability and robot repeatability \cite{gillespie2001general}. In \textit{Cooking with Robots}, Sugiura \etal proposed collaborative mobile counter-top robots to assist with stove cooking. They found that collaboration is more practically feasible and encourages more natural human-robot interactions, but reported manipulation limitations \cite{sugiura2010cooking}. In the decade since, few collaborative cooking breakthroughs have been reported.

This paper presents a new approach to the collaborative cooking problem. To achieve feasibility in a dynamic environment, we propose that the system should instead take a more passive role, assisting user-led actions rather than replacing the chef. Removing the robot arm avoids manipulation challenges with currently available technology. We instead leverage computer vision techniques to inform contextual instructions for the user to execute (Figure~1). We introduce a system that balances manual manipulation with automatic heat control, promoting natural and dynamic interactions. Our main contributions include: 

\begin{itemize}
  \item A first-of-its-type human-centered approach to computational cooking.
  \item A novel system for recipe image collection, training, and application of computer vision.
  \item An open source implementation of the hardware and software, to enable the community to build image datasets and accelerate progress by developing ideas in the open.
\end{itemize}

This system is proposed as a key enabling technology on the critical path to fully autonomous (non-passive) systems of the future. 

\section{Related Work}

\subsection{Cooking Robotics}
The semi-structured kitchen environment poses interesting challenges for robotics research --- perception in cluttered environments, object manipulation, and complex cooking actions --- while allowing for simplification due to consistency of kitchen spaces, tools, tasks, and recipe structure \cite{bollini2013interpreting}. Recipe websites have been parsed into low-level action sequences \cite{bollini2013interpreting, inagawa2020japanese}. Cooking videos on YouTube have been used to automatically learn and execute action sequences \cite{zhang2019robot}.

High degree-of-freedom robot arms \cite{bollini2013interpreting, junge2020improving, inagawa2020japanese} and humanoid robots \cite{gravot2006cooking, zhai2012kinematic, yi2020task} are typically used to mimic human-kitchen interaction in robotics research. Such studies do not allow for shared human-robot workspaces, limiting the real-world usefulness of the methods. Moley \cite{moley2020} motion-captures and replicates recipes demonstrated by a user, but cannot collaborate on the cooking process. In \textit{Cooking with Robots}, Sugiura \etal proposed Cooky, a human-in-the-loop stove-top cooking system with mobile robots.  Cooky predates modern computer vision techniques, so users must attach visual markers (paper tags) to utensils and ingredients before before cooking \cite{sugiura2010cooking}. The robots allow for collaboration in a shared workspace but only provide rudimentary cooking capabilities. 

\subsection{Digital Kitchen Interfaces}
Tablets and other screens are an increasingly important component of the kitchen. Voice assistance has been used to reduce the need to use screens with dirty hands \cite{kumagai2017cooking}. Julia is a commercially available device that combines various preparation and cooking functions with a tablet interface and voice assistant \cite{julia2020}. With ChopTop, Celik \etal explore an interactive chopping board that combines simplified on-screen recipe instructions with weighing and timing functionality for inexperienced cooks \cite{celik2018choptop}. Augmented Reality has been explored as a solution for displaying guidance based on actions in the kitchen \cite{miyawaki2008virtual, d2018augmented}. The complementary aspects of the above interfaces could feasibly be integrated into a collaborative computational cooking system, benefiting from supplementary information from computer vision.

\subsection{Computer Vision}
Food images have many complex features that are challenging to define, posing significant computational challenges to conventional image based decision making approaches \cite{mezgec2017nutrinet}. Deep learning has outperformed manual feature extraction and conventional machine learning techniques in food quality analysis \cite{zhou2019application}. Deep learning has been used in food classification, quality detection, calorie estimation and food contamination investigation \cite{fellows2009food, bhotmange2011application, zhou2019application}.

\begin{figure}[h]
  \centering
  \includegraphics[width=\linewidth]{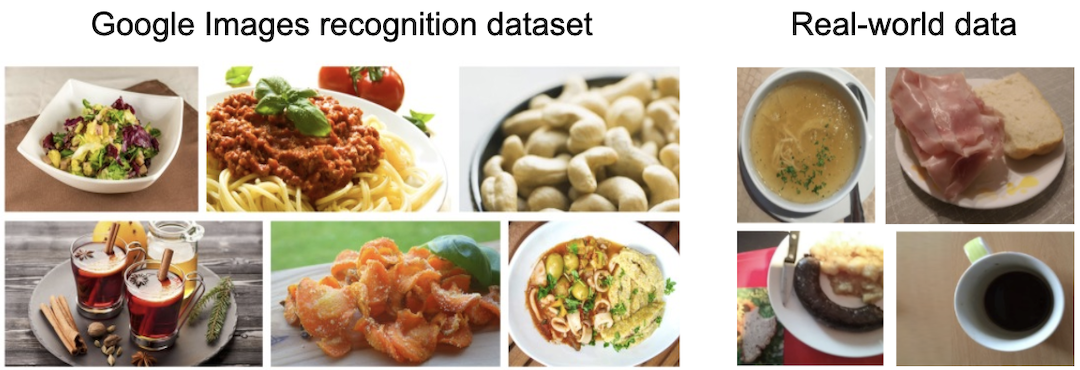}
  \caption{NutriNet Dataset: Real-world smartphone images differ significantly in style and composition to training images scraped from Google Images, causing accuracy losses. Adapted from \cite{mezgec2017nutrinet}.}
  \Description{A selection of photographs of food and drink. On the left: 6 photos with the title `Google Images recognition dataset'. The photos are bright, professional quality 'stock images' of beautiful dishes. On the right: 4 photos with the `title Real-world data'. The photos are poor camera quality, dark, poorly composed images of uninteresting dishes.}
  \label{nutrinet}
\end{figure}

A large training dataset is critical to food image recognition as it enables learning of general features, reducing the effects of overfitting \cite{mezgec2017nutrinet}. NutriNet \cite{mezgec2017nutrinet} introduces an architecture trained on 220,000 images gathered using Google Images searches of food and drink items. While 94.5\% top-five accuracy was achieved on the testing subset, top-five accuracy for real-world images was 55\%. The authors cite noise, occlusion, and overfitting to the Google Images dataset (Figure~\ref{nutrinet}) as possible causes \cite{mezgec2017nutrinet}. 

The Recipe1M \cite{salvador2017learning} and Recipe1M+ \cite{marin2019recipe1m} datasets include 1 million internet recipes and their corresponding completed dish images. The Cookpad dataset \cite{harashima2017cookpad} (1.6 million images and recipes) also includes 3.1 million cooking \textit{process} images collected by the Cookpad recipe app. Ingredient \textit{state} is important to determining subsequently performed actions (whether a tomato is whole, sliced, or diced for example). A food state classifier was trained to identify 11 different states for 17 ingredients \cite{jelodar2018identifying}. The available datasets either include images upon recipe completion or at a few key points. Images captured at regular time intervals throughout the entire cooking process may be useful for process control. To the best of our knowledge, no \textit{`continuous cooking process datasets'} have been established. 

MISO Robotics have developed a device that combines a camera with a far-infrared thermal sensor data to assist grilling and frying tasks \cite{zito2020robotic, sinnet2018multi}. The MISO AI computer vision model tracks burger cooking progress, increasing operator efficiency and reducing failure rates \cite{miso2020ai}.

\section{Proposed architecture}

Desirable characteristics for a collaborative cooking system are as follows:
\begin{enumerate}

    \item{\textit{Dynamic:}}
    The system should operate in a human-centric workspace and adapt to interventions from the chef. Fast responses to interventions should be a characteristic of the system; frame rate will ideally be maximized and latency minimised. 

    \item{\textit{Small:}} 
    Ensuring the system fits comfortably on a kitchen counter is a rational approach.

    \item{\textit{Accurate:}}
    The system should detect and control cooking temperature to an approximately human level. 

    \item{\textit{Robust:}}
    Food computer vision attempts in the literature have reported limited real-world success. Perception robustness is desired; training on visually consistent in-context images should aid model accuracy.

\end{enumerate}
Desirable characteristics \textit{motivated by the need for large image datasets} are as follows:

\begin{enumerate}
    \setcounter{enumi}{4}

    \item{\textit{Automated labelling:}}
    Suitable cooking process image sets do not exist in the literature; an interface should facilitate easy, real-time image labelling. Researchers should be able to generate training sets for their own use cases (recipes).  
    
    \item{\textit{Open source:}}
    An open source community could crowd-source labelled images and share recipe models. The system should utilise of off-the-shelf and 3D-printed components. 
    
    \item{\textit{Accessible:}}
    The system should simplify the model training process, making it accessible to cross-disciplinary researchers without machine learning expertise. An API should facilitate development of further functionality.
\end{enumerate}

\begin{figure*}[h]
  \centering
  \includegraphics[width=\linewidth]{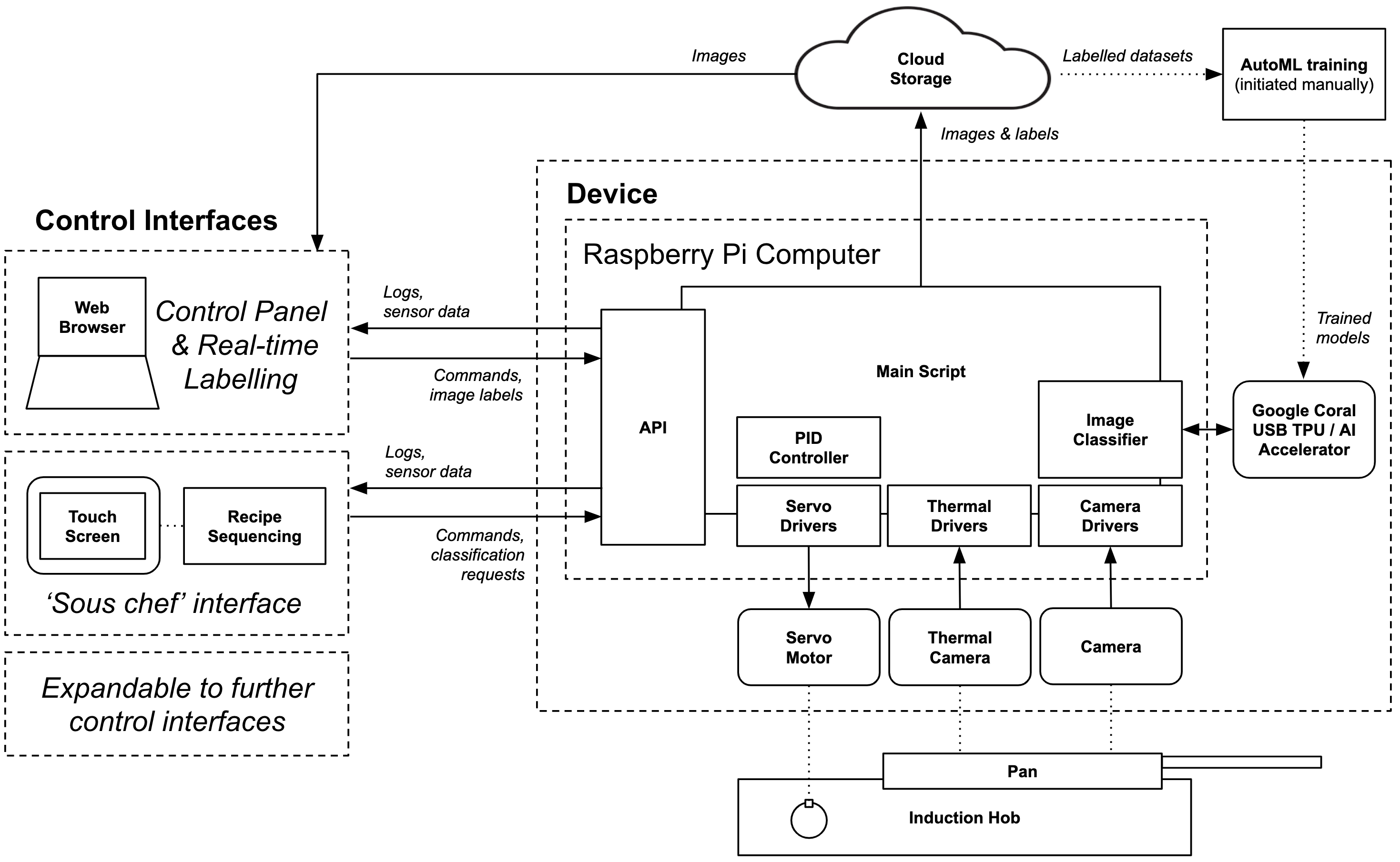}
  \caption{Computational Cooking System Overview. The API facilitates expansion to new modules and control interfaces.}
  \Description{A system diagram showing 1 main 'device' module and several peripheral modules. The 'device' module contains a Raspberry Pi Computer, Servo Motor, Thermal Camera, Camera, and Google Coral USB TPU AI accelerator. Below, the servo, thermal and camera scripts connect to a pan and induction hob. Within the Raspberry Pi, boxes represent scripts that connect to components, including Servo, Thermal, and Camera drivers. A PID controller script connects to the servo driver. An image classifier script connects to the Coral AI accelerator. Finally, An API script connects to the peripheral modules called Control Interfaces including 1) Control Panel and Real-time Labelling web browser and 2) a 'Sous-chef' interface, recipe sequencing script and touchscreen. Logs and sensor data exit the API and commands enter the API. Above, a cloud storage module receives images from the Raspberry Pi main script, sending images to the Control interfaces and sending labelled datasets to a AutoML training module, which sends trained models to the Google Coral AI accelerator.}
  \label{system}
\end{figure*}

Figure~\ref{system} shows an overview of our system. We add sensing and control functionality to a commercially available induction cooker. Researchers can use the Control Panel and Real-time Labelling interface (Section 4.4) for development, and a chef would use the `Sous chef' assistant interface while cooking. The following section describes the implementation of our new system designed to satisfy the above design considerations. 

\section{Implementation}

\subsection{Sensing}
\begin{figure}[h]
  \centering
  \includegraphics[width=\linewidth]{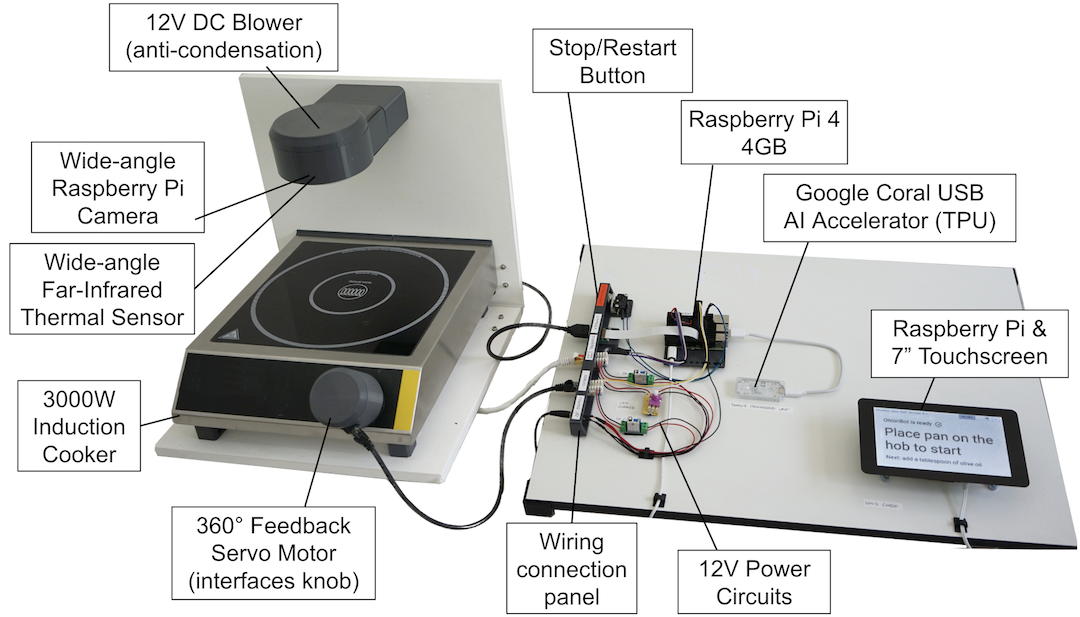}
  \caption{System hardware overview.}
  \Description{Labelled boxes point to the device described in Figure 1, on a blank white background. The upper module includes a 12V DC blower (anti-condensation), a Wide-Angle Raspberry Pi camera and a Wide-Angle Far-Infrared Thermal sensor. The stove is labelled as a 3000W induction cooker. Mounted to the front is a module containing a 360 degree feedback servo motor that interfaces the hob. These modules are wired to the components board, which includes a stop/restart button, a Raspberry Pi 4GB, a Google Coral USB AI accelerator, a Raspberry Pi & 7 inch touchscreen, 12v power circuits and a wiring connection panel. The screen says 'Place a Pan on the Hob to Start' }
  \label{hardware}
\end{figure}

A 120\degree \ wide-angle lens enables camera placement close to the pan to minimize overall height (Figure~\ref{hardware}). Image classification is performed on the edge (on device); initial testing suggested that inferencing would slow frame rate, so a Google Coral AI USB Accelerator (Tensor Processing Unit) improves performance. We use a MLX90640 Far-Infrared Sensor Array to measure pan temperature, an established method in the literature \cite{sinnet2018multi, vollmer2017infrared}. Components are housed in a 3D-printed PLA enclosure, for which 3D files are publicly available. 

\begin{figure*}[h]
  \centering
  \includegraphics[width=\linewidth]{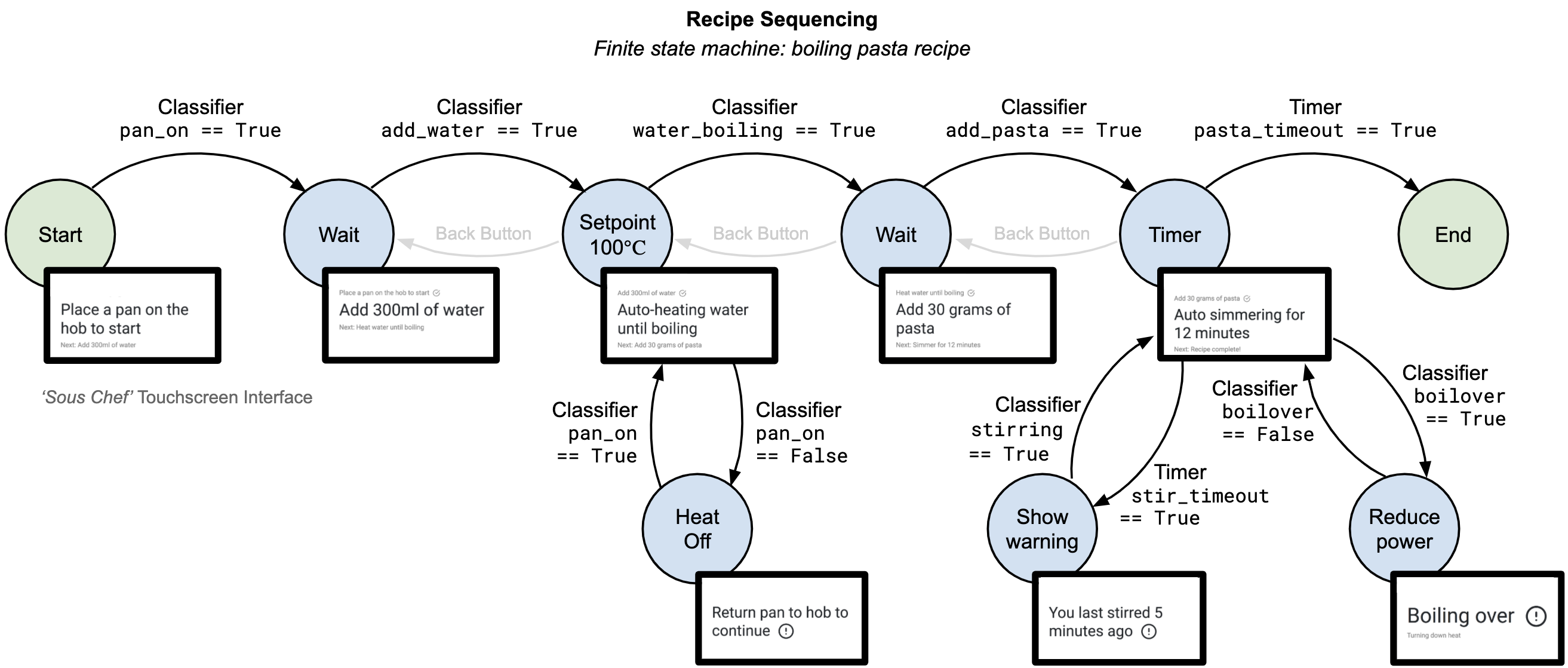}
  \caption{An example Finite State Machine for boiling pasta. The `sous-chef' interface displays recipe guidance (shown) according to pan conditions. This simple recipe can be followed in its entirety without touching a screen.}
  \Description{A finite state machine (FSM) diagram with the title Recipe Sequencing / Finite state machine: boiling pasta recipe. Circles show a start and end point. A visualisation of the touchscreen overlays each of the FSM state circles. The flowchart has the following elements, screen described in brackets:
  Main elements: 
  1. Start (Place a pan on the hob to start). Go to 2 when Classifier pan_on == True. 
  2. Wait (Add 300ml of water). Go to 3 when Classifier add_water == True. 
  3. Setpoint 100C (Auto heating water until boiling ). Go to 4 when Classifier water_boiling == True. 
  4. Wait (Add 30 grams of pasta). Go to 5 when Classifier add_pasta == True. 
  5. Timer (Auto simmering for 12 minutes. Go to 6 when Timer pasta_timeout == True. 
  6. End 
  Other elements:
  7. Enter from 3 when Classifier pan_on == False. Heat off (return pan to hob to continue). Exit to 3 when Classifier pan_on == True.
  8. Enter from 5 when Timer stir_timeout == True. Show warning (you last stirred 5 minutes ago). Exit to 3 when Classifier stirring == True.
  9. Enter from 5 when Classifier boilover == True. Reduce power (Boiling over !). Exit to 3 when Classifier boilover == False.
  }
  \label{finitestate}
\end{figure*}

\subsection{Control}
We use a Buffalo 3000W Induction Cooker, controlled by a front-mounted servo motor. A Parallax Feedback 360\degree \ Servo provides appropriate range of motion and returns angular position with an internal Hall effect sensor. We use a PID  controller to manage temperature, as established in the literature \cite{visioli2006practical}. A Raspberry Pi Foundation 7” Touchscreen display provides instructions, reminders and warnings. While actuation is limited to induction power control, the human-machine-interface provides a powerful secondary control method that leverages the dexterity and flexibility of the human as an actuator.

Recipes are programmed following the finite state machine format presented in Figure~\ref{finitestate}. This example shows a variety of unique assistive functionality enabled by image classification techniques and temperature control. For this simple pasta boiling example, the device can:
\begin{itemize}
    \item Automatically advance through recipe instructions; no need to interact with a screen with messy hands.
    \item Automatically switch off the power on pan removal, then switch it back on on pan return.
    \item Hold the pan at a particular temperature in degrees.
    \item Automatically start a timer for a recipe instruction, then turn off the heat when the timer is complete.
    \item Show a reminder when the pan has not been stirred for a duration, then hide the warning when the pan is stirred. 
    \item Show a warning if the pan boils over, and turn down the heat.
\end{itemize}
These functions aim to reduce the cognitive load on the chef, allowing their attention to be focused on the more creative aspects of cooking. We hope that by facilitating perfect execution of recipes every time we will enable convenience, waste-reduction, cost, and health benefits for users of all ability levels.

\subsection{Computer Vision}
The unstructured kitchen workspace is a challenging environment for computer vision \cite{bollini2013interpreting}. We propose a fixed field-of-view including only the stovetop, creating a more structured environment to improve perception robustness. Fixed LED lighting above the pan further improves image consistency. 

The hot pan creates tough conditions for the camera and thermal array. Initial testing revealed that image quality was deteriorated due to condensation on the sensors. We developed a mechanism using a Sunon 12V DC Brushless Blower to flow a layer of dry air over the sensors to minimise condensation and temperature variation. 

Studies have shown poor real-world results in training general food-classification and general food-state-classification models \cite{liu2016deepfood, mezgec2017nutrinet, jelodar2018identifying}. However, \textit{cooking} image classification can be simplified by training individual models for individual recipes. We realized that we only need to identify key events at which actions are to be performed. We establish a \textit{milestone-based} approach, shown in Figure~\ref{milestones}. We classify key events for a single recipe, dramatically simplifying the classification problem. For example, we do not attempt to learn the features of an onion in any context; we only differentiate the \textit{`Onions added to the pan'} milestone from the \textit{`Onions cooked until soft'} milestone, so the appropriate action may be taken at each step. 

\subsection{Image Labelling}
\begin{figure}[h]
  \centering
  \includegraphics[width=1\columnwidth]{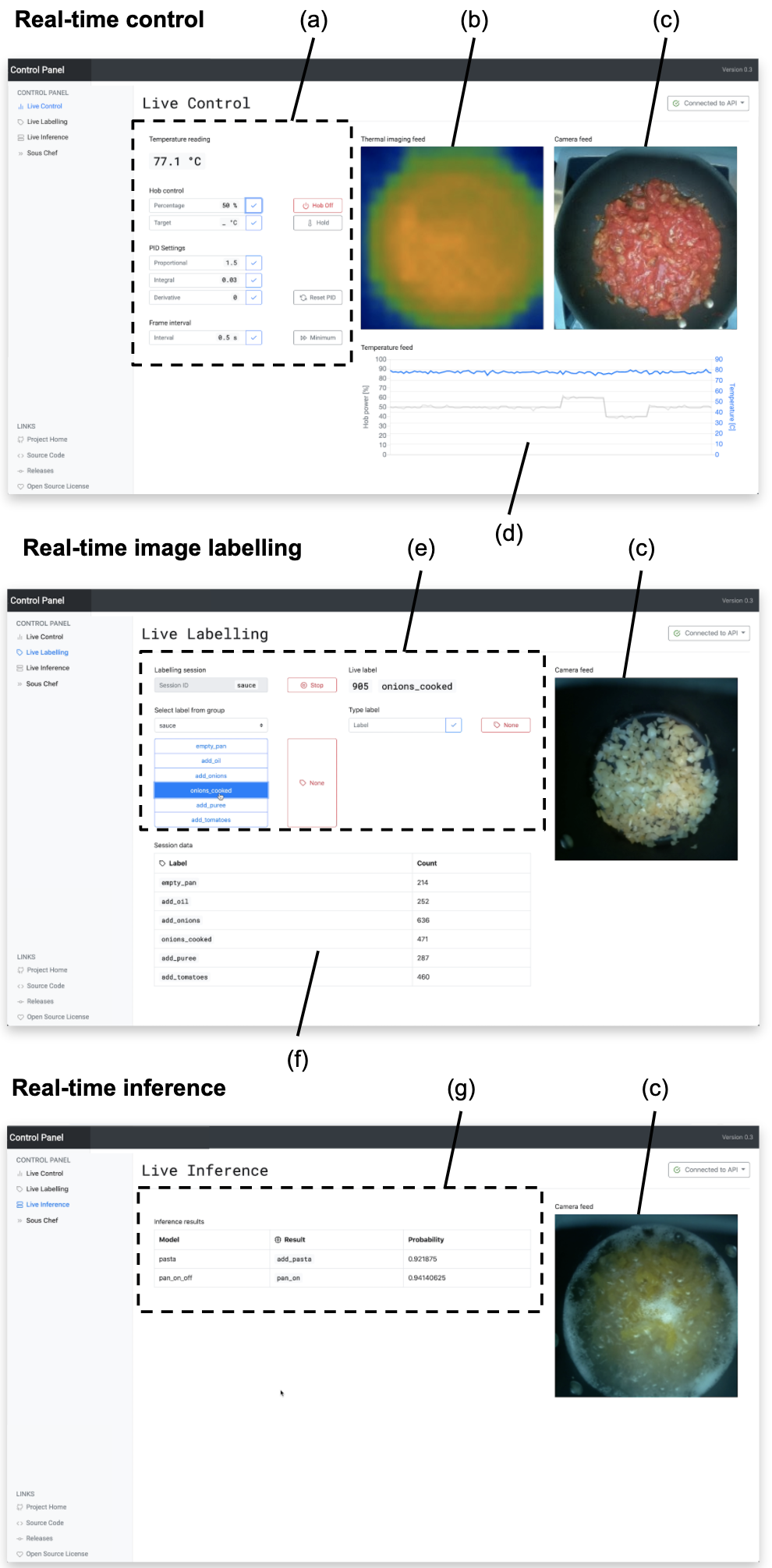}
  \caption{The Control Panel interface for researchers connects to the API. It includes: (a) parameter adjustment, (b) thermal array live stream, (c) camera live stream, (d) time series temperature plot, (e) real-time labelling controls, (f) image set information, (g) image inference information.}
  \Description{3 web interface screenshots with the titles (1) Real-time control, (2) Real-time image labelling and (3) Real-time inference. REAL TIME CONTROL includes (a) parameter adjustment web form inputs, (b) thermal array freeze frame of round hot pan, (c) camera freeze frame of tomato sauce, (d) arbritrary time series temperature plot. REAL TIME IMAGE LABELLING includes (c) camera freeze frame of onions (e) web form controls for labelling, (f) table of labels collected so far. REAL TIME INFERENCE includes (c) camera freeze frame of pasta, (g) table of current inference data about pasta}
  \label{controlpanel}
\end{figure}

Suitable milestone-based image datasets do not exist in the literature, so it is necessary for the system to self-generate data for each recipe. We designed a web interface (Figure~\ref{controlpanel}) for labelling images in real-time while cooking. Using the Control Panel, researchers can monitor and control all elements of the system. Labelled images are continuously uploaded to Google Cloud at 0.5 second intervals, and a database of labels is automatically created, formatted for import into AutoML.

\begin{figure}[h]
  \centering
  \includegraphics[width=0.9\columnwidth]{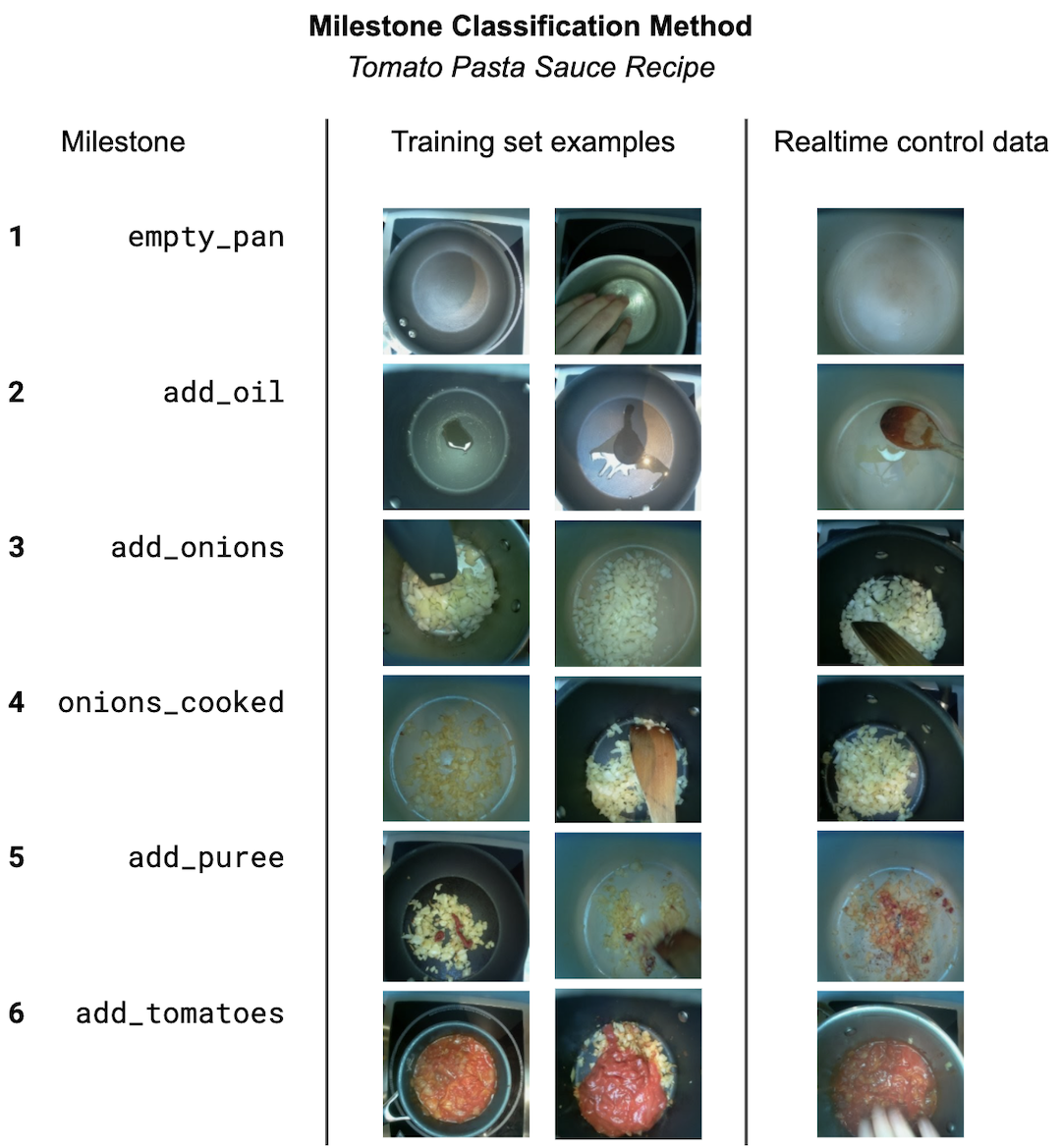}
  \caption{Milestone classification method, showing image classes for each key milestone in the recipe. Contextual noise and variation are manually introduced to every class. As training images are context, real-time control images are shown to be consistent and accurate to the training set.}
  \Description{Table entitled Milestone classification method: tomato pasta sauce recipe. There are 3 columns (1) Milestone, (2) Training set examples, (3) Realtime control data. The milestone column includes 6 labels: empty_pan, add_oil, add_onions, onions_cooked, add_puree, and add_tomatoes. The training set column shows top down pan view images of these recipe steps. The realtime control data column shows visually consistent images of each milestone, suggesting that training images match realtime images. 
}
  \label{milestones}
\end{figure}

We generated image datasets for the pasta and tomato sauce test recipes (from Figures~\ref{finitestate} and \ref{milestones}) using the system. We initially found that classes were unbalanced due to variation in milestone duration, so slower milestones were undersampled to balance classes. Over 5 cooking sessions we captured 2,045 images across 4 milestones for pasta, and 4,042 images across 6 milestones for tomato sauce.

\subsection{Model Training}
To simplify the model training process for cross-disciplinary researchers, we take advantage of Google's AutoML automated training service. Google does not disclose architecture information, but we speculate that training is based on established architectures such as ResNet \cite{he2016deep}, likely alongside transfer learning techniques \cite{wong2018transfer}. 

Unlike scraped \cite{mezgec2017nutrinet, liu2016deepfood, salvador2017learning, marin2019recipe1m} or crowd-sourced \cite{harashima2017cookpad} datasets in the literature, our training images are captured on-device. As such, we should not expect overfitting to training data due to contextual information such as style and composition. However, we did find that models will associate stirring utensils with a particular milestone (always stirring onions with a wooden spoon, for example, causes confusion when a wooden spoon appears in a different milestone). As such, we added more examples of contextual noise to all classes of the training data, including variation in pans, lighting, and stirring utensils. Models were trained using Google AutoML until performance no longer improved (typically 1-2 node-hours). At 0.5 confidence threshold, AutoML achieved 100\% test set precision for pasta milestone classification and 99.5\% test set precision for tomato sauce milestone classification. The confidence threshold was also set at 0.5 for real-world testing, with iterations performed until each milestone triggered at the correct time. 

\section{Experimental validation}
The full system has been shown to work consistently. We performed preliminary tests with a student in our lab with no knowledge of the system. The student must cook pasta and tomato sauce following the on-screen instructions exactly. This simple recipe requires the system to successfully demonstrate 9 classifier-informed `proceed to next step' events, 3 time-based events, and 3 classifier-informed `warning' interventions. The most challenging vision task is detecting when frying onions become soft. The test is failed if the system advances too early or late, so the skip-forward, skip-back or cancel-warning buttons are pressed. The system was shown to work repeatably over 10 initial tests of approximately 30 minutes each, and the test user was able to reproduce these results. 

\section{Open Source Collaborative Computational Cooking}
We present here key enabling components for collaborative computational cooking; outlining feasible approaches to connected systems and computer vision methods. The critical remaining component is data. Food brings people together; we hope that an open-source computational cooking movement could inspire a community to build their own systems. A first-of-its-kind \textit{`continuous cooking process database'} could be crowd-sourced, with users labelling and sharing their images and recipes. Metadata captured by the system is also incredibly rich: recipes, temperatures, ingredients, and corrective human inputs are recorded. A large, rich image dataset would enable new, advanced research with deep learning. 

We open source our work in aid of accelerating research progress by developing in the open. We provide an implementation of our system including: CAD files, bill of materials (BOM), real-time labelling interface, device control panel interface, and `sous chef' touchscreen interface\footnote{Available at \href{https://github.com/onionbot}{github.com/onionbot}.}. The required components, excluding the 3D-printed enclosure, are available off-the-shelf for under \$500.

\section{Discussion and Limitations}
The test user successfully cooked a meal with the system, demonstrating 9 successful semi-autonomous interventions. Our user-led assistance approach to cooking automation allowed the system to operate in the dynamic kitchen environment. This shows obvious size and feasibility advantages over previous solutions using robot arms in static-only environments \cite{bollini2013interpreting, junge2020improving, inagawa2020japanese, yi2020task, gravot2006cooking, zhai2012kinematic}. Incremental BOM cost for an appliance manufacturer incorporating this technology into an existing product would be a fraction of the cost of any robot arm.

However, end users may expect a more autonomous system; we require users to execute many cooking actions themselves. Notwithstanding, we believe that the value of the system is instead demonstrated by its capability to reduce cognitive load on the chef by acting as a `second pair of eyes'. We argue that intelligent screen-based guidance can provide meaningful assistance and encourages more natural human-robot collaboration. We have not yet established whether the system can add value for more complex recipes, which we see an opportunity for future work.

Vision system responsiveness depends on frame rate. We found that our 2 Hz frame rate gives an instantaneous event response time of approximately 2 seconds (propagation time for the rolling average filter). Response time to more gradual changes depends on confidence. Future work will improve responsiveness through frame rate optimisations.   

PID control proved to be an adequately accurate temperature control method. We do not explore how descriptive vision outputs such as \textit{`pasta is boiling over’} should feed into precise PID control systems. Future work could explore Fuzzy Logic Control Systems as an alternative. Fuzzy logic, also used in the food industry, mimics human reasoning in allowing computers to behave imprecisely, enabling descriptive inputs to control precise systems \cite{fellows2009food}.

For computer vision robustness, examples of variation and noise must be manually introduced to every class of training data for each recipe. This is a laborious process. Future work may explore Transfer Learning, where knowledge gained from solving one problem is applied to new problems \cite{goodfellow2016deep}. Robustness information such as \textit{‘ignore features that look like hands’} could be transferred so that examples are not necessary in new training sets. 

Images were successfully labelled in real-time using our control and labelling software. A limitation of the `milestones' classification approach is that individual models must be trained for any new recipe. However, in future this process could be automated. 

Recipes must be manually encoded into sequences of functions corresponding to trained models. Studies have explored automatic parsing of recipe websites into low-level action sequences \cite{bollini2013interpreting, inagawa2020japanese}. Our recipe sequencing structure could feasibly be generated by Natural Language Processing techniques in the literature \cite{goodfellow2016deep}.

To the best of our knowledge, `\textit{continuous cooking process datasets}' have not yet been established in the literature. Networked devices with basic functionality could incentivize users to start sharing labelled images and recipes. A large, rich, crowd-sourced cooking dataset could facilitate development of advanced deep learning models, with new functionality available as over-the-air updates. 

\section{Conclusion}
This paper proposes a novel system for assisted cooking in dynamic environments. We show technical feasibility for our system, confirming that it can successfully assist a test user in executing a simple meal. The system uses computer vision techniques and temperature control to provide intelligent and natural interventions during stove-top cooking. Users cook with the support of instructions and guidance from the digital \textit{Sous Chef}. We argue that computational cooking systems should take predominantly passive roles until challenges with current robots in dynamic workspaces are overcome. In the meantime, these assistive devices should build image databases and refine vision models for future automation. Finally, we emphasize that large recipe image datasets will be critical to any cooking automation system. We hope that developing systems and datasets in the open will accelerate the progress of computational cooking. We provide an open source implementation of our work to encourage future research. 

\newpage

\bibliographystyle{ACM-Reference-Format}
\balance
\bibliography{references.bib}

\end{document}